\documentclass{webofc}
\usepackage[varg]{txfonts}   

\newcommand{\eq}[1]{\begin{align} #1 \end{align}}

\begin{document}
\title{Fluctuations in heavy ion collisions\\ and global conservation effects}

\author{\firstname{Roman} \lastname{Poberezhnyuk}\inst{1,2}\fnsep\thanks{\email{rpoberezhnyuk@bitp.kiev.ua, poberezhnyuk@fias.uni-frankfurt.de}} \and
        \firstname{Volodymyr} \lastname{Vovchenko}\inst{3,4,2} \and
        \firstname{Oleh} \lastname{Savchuk}\inst{1,5,2} \and
        \firstname{Volker} \lastname{Koch}\inst{4} \and
        \firstname{Mark}~\lastname{Gorenstein}\inst{1,2} \and
        \firstname{Horst} \lastname{Stoecker}\inst{2,6,5}
}

\institute{
Bogolyubov Institute for Theoretical Physics, 03680 Kiev, Ukraine
\and
           Frankfurt Institute for Advanced Studies,
D-60438 Frankfurt am Main, Germany
\and
           Institute for Nuclear Theory, University of Washington, Box 351550, Seattle, WA 98195, USA
\and
        Lawrence Berkeley National Laboratory, 1 Cyclotron Road,  Berkeley, CA 94720, USA
\and
        GSI Helmholtzzentrum f\"ur Schwerionenforschung GmbH, Planckstr. 1, D-64291 Darmstadt, Germany
\and
        Institut f\"{u}r Theoretische Physik, Goethe Universit\"{a}t Frankfurt, D-60438 Frankfurt am Main, Germany
}

\abstract{%
Subensemble is a type of statistical ensemble which is the generalization of grand canonical and canonical ensembles.
The subensemble acceptance method (SAM) provides general formulas to correct the cumulants of distributions in heavy-ion collisions for the global conservation of all QCD charges. The method is applicable for an arbitrary equation of state and sufficiently large systems, such as those created in central collisions of heavy ions.
The new fluctuation measures insensitive to global conservation effects are presented.
The main results are illustrated in the hadron resonance gas and van der Waals fluid frameworks.
}

\maketitle
\section{Introduction}
\label{intro}

Fluctuations are essential in studies of the quantum chromodynamics (QCD) phase diagram in first-principle lattice QCD simulations and heavy-ion collision (HIC) experiments.
They are sensitive to fine details of interactions, for example to a phase structure including searched-for phase transition and a critical point (CP) of QCD matter~\cite{Stephanov:1998dy,Stephanov:1999zu,Athanasiou:2010kw,Stephanov:2008qz,Vovchenko:2015pya,Poberezhnyuk:2019pxs}.
Fluctuations are characterized by cumulants $\kappa_n$.

In QCD simulations and statistical models, the grand canonical ensemble (GCE) is mostly used.
In GCE, cumulants of quantity $B$ distribution can be calculated by taking the derivative of pressure $p$ over the corresponding chemical potentials $\mu_B$:
\eq{\label{gce-limit}
\kappa_n[B]=V T^3 \chi^B_n,~~~~~\chi^B_n\equiv \partial^n(p/T^4)/\partial(\mu_B/T)^n.
}

GCE implies that the total system with volume $V$ is separated in the coordinate space into the observed subsystem with volume $V_1$ and the rest of the system, which is called reservoir (or thermostat) with volume $V_2$, $V=V_1+V_2$.
The subsystem and thermostat can freely exchange particles (or charge $B$).
In the total system, the charge is fixed, and fluctuations are absent.
However, the charge fluctuates within the subsystem, and we measure fluctuations in this subsystem.

For the observed subsystem to be in GCE, the following requirements must be satisfied:
Firstly, $V_1$ must be much larger than the correlation length, $V_1\gg \xi$,  i.e., thermodynamic limit is realized.
Secondly, the subsystem must be a small part of the total system, $V\gg V_1$. In this case, the effects of global charge conservation can be neglected.

The size of the created system in HIC is not large enough to satisfy these two requirements simultaneously.
Thus, the results of experimental measurements of cumulants can't be compared directly with grand-canonical predictions
\footnote{
Other technical issues present in the experiment also hinder the direct comparison. 
Here we focus only on the effects of global conservation of QCD charges.
}
and
the global charge conservation effects
must be taken into account.
\footnote{Note that in HIC, event-by-event fluctuations are measured in momentum space rather than coordinate space.
Therefore, for the method's applicability, the momentum-space correlations must be substantial due to collective flow.
We claim that the strength of space-momentum correlations is sufficient, at least at high collision energies of the Large Hadron Collider, LHC. 
For the estimation of effects of thermal smearing and imperfect space-momentum correlations, see Ref.~\cite{Vovchenko:2020kwg}.}
The developed subensemble acceptance method (AM)~\cite{Vovchenko:2020tsr, Poberezhnyuk:2020ayn,Vovchenko:2020gne} allows to account for global conservation effects.

\section{Subensemble Acceptance Method}
\label{sam}

Subensemble is a new type of statistical ensemble, a generalization of GCE. It corresponds to a finite subsystem in contact with a finite thermostat.
We partition a thermal system with fixed conserved charge $B$ and temperature $T$ into two subsystems.
\footnote{Assuming the subsystem is large, we neglect surface effects.} The Hamiltonian $H$ can be written as $H=H_1+H_2$.
Then the canonical partition function of the total system can be written as 
\eq{
Z(T,V,B)
  = \sum_{B_1} Z
  (T,\alpha V,B_{1}) Z
  (T,\beta V,B-B_{1}),~~~~~\alpha\equiv V_1/V,~~~~~\beta\equiv 1-\alpha. \label{ZB}
}

We express cumulants $\kappa^B_n$ of fluctuations inside the subsystem through the corresponding grand canonical susceptibilities, $\chi^B_n\equiv \partial^n(p/T^4)/\partial(\mu_B/T)^n$. This can be done in the thermodynamic limit~\cite{Vovchenko:2020tsr}, yielding:
\eq{\nonumber
\kappa_1[B_1] &= \alpha VT^3 \,  \chi_1^B,~~~~~
\kappa_2[B_1]  = \alpha VT^3 \, \beta \chi_2^B,~~~~~
\kappa_3[B_1]  = \alpha VT^3 \, \beta (1-2\alpha) \chi_3^B, \\
\label{eq:kappa1B}
\kappa_4[B_1]  &= \alpha VT^3 \, \beta \left[ 
 \, (1-3\alpha \beta) \, \chi_4^B
- 3 \, \alpha \, \beta \, \frac{(\chi_3^B)^2}{\chi_2^B}
\right].
}

The model-independent formulas \eqref{eq:kappa1B} provide corrections for global charge conservation.
In the case $\alpha\to 0$, the charge conservation effects are absent, and the cumulants $\kappa_n$ reduce to their GCE limit~\eqref{gce-limit}. In the opposite case $\alpha\to 1$, cumulants approach the limit of the canonical ensemble where fluctuations are absent. 

See Ref.~\cite{Vovchenko:2020tsr} for results on cumulants up to 6th order, as well as on special cumulant ratios: scaled variance $\omega\equiv\kappa_2/\kappa_1$, skewness $S\sigma\equiv\kappa_3/\kappa_2$, and kurtosis $\kappa\sigma^2\equiv\kappa_4/\kappa_2$.

\section{Applications and Extensions}
\label{applications}

\begin{figure}[h]
\centering
\includegraphics[width=0.49\textwidth]{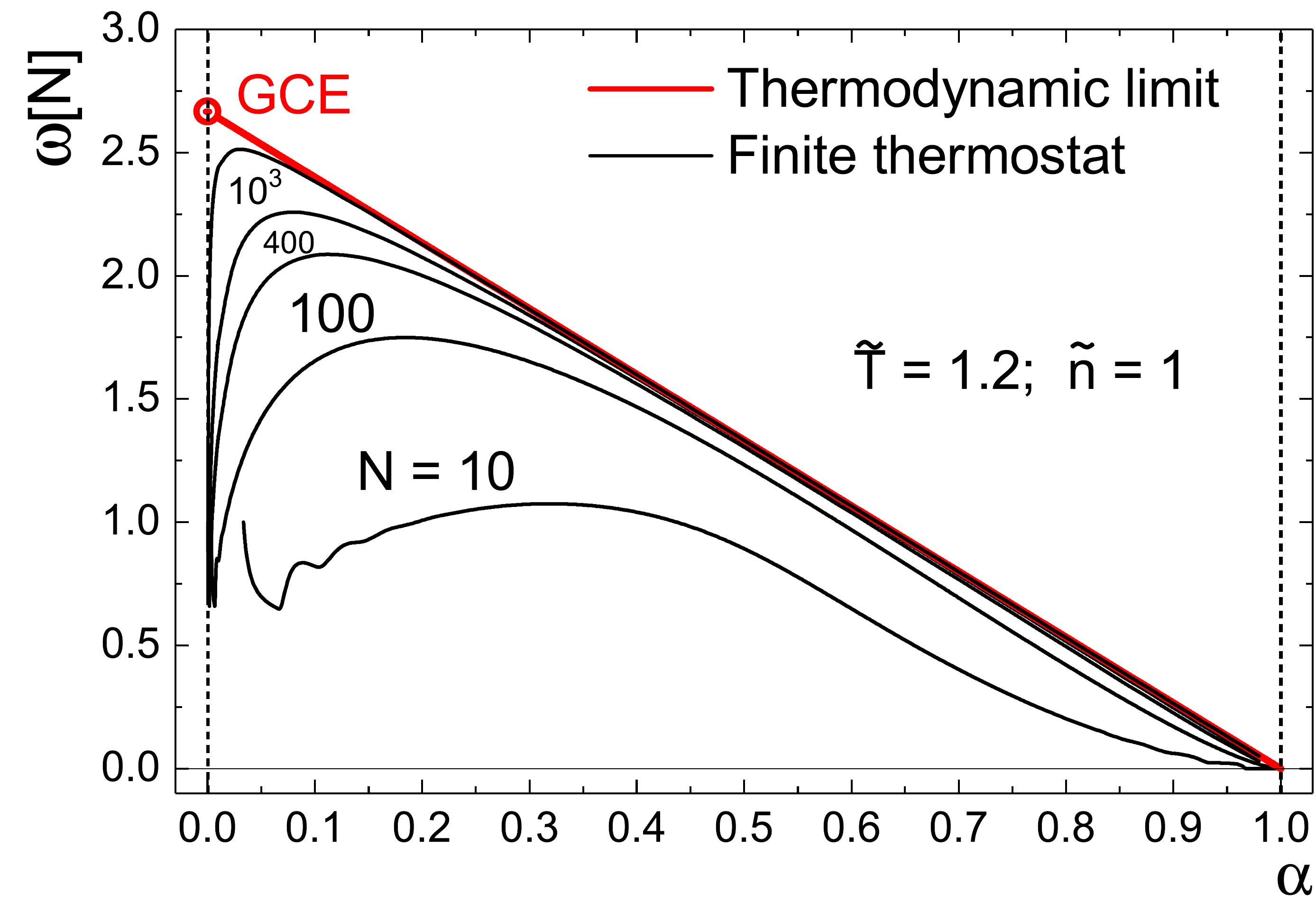}
\includegraphics[width=0.49\textwidth]{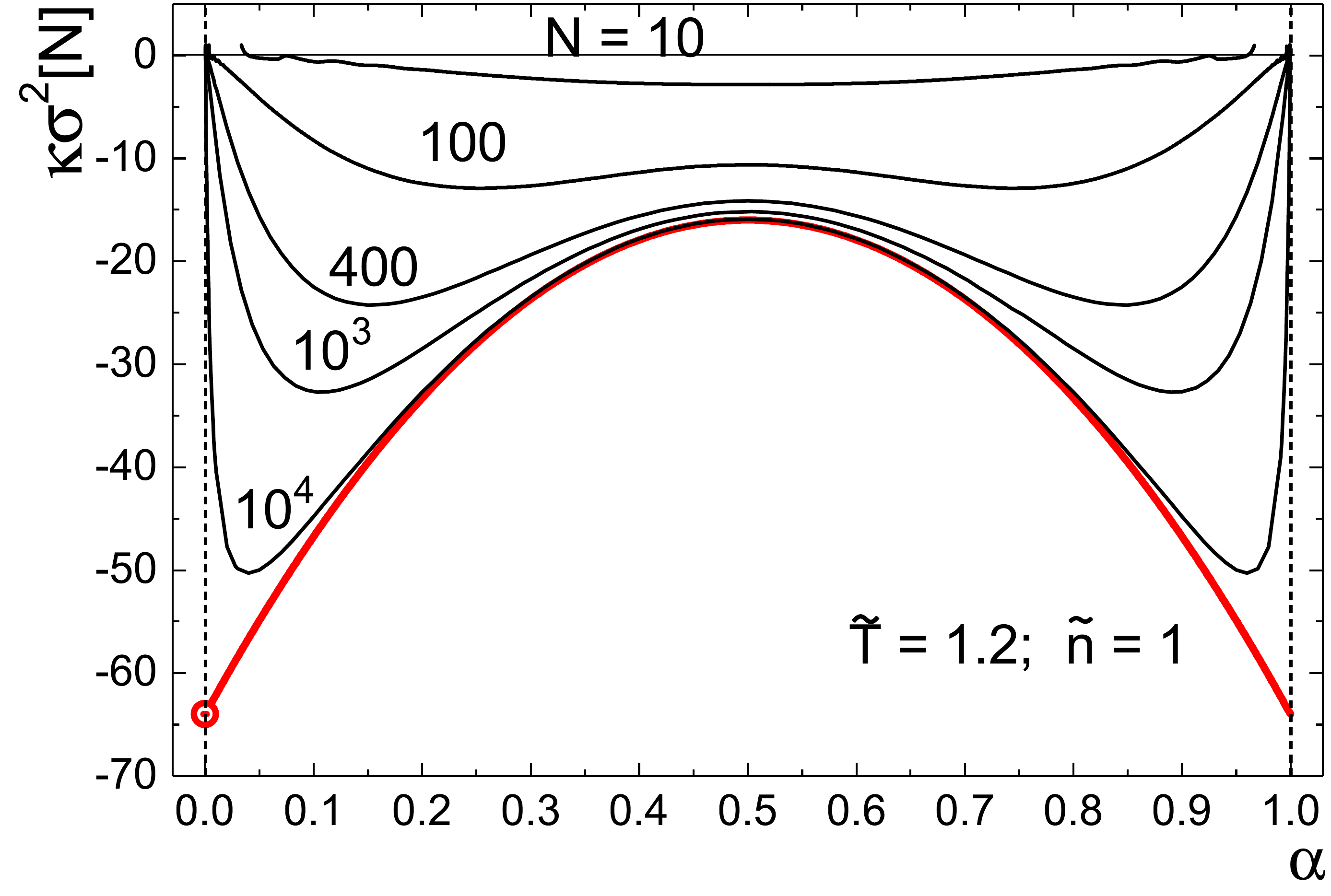}
\caption{Scaled variance (left) and kurtosis (right) of particle number (conserved charge) fluctuations as functions of the parameter $\alpha$. The red lines are obtained from the SAM formulas~\eqref{eq:kappa1B}. The black lines represent the numeric results calculated explicitly from the vdW partition function~\cite{GNS} using Eq.~\eqref{ZB} for different values of the total number of particles in the system $N_0$. The red dots at $\alpha=0$ represent GCE values~\cite{Vovchenko:2015uda}.
Taken from~\cite{Poberezhnyuk:2020ayn}.
}
\label{fig-1}       
\end{figure}

The subensemble acceptance analytic formulas have been checked~\cite{Poberezhnyuk:2020ayn} for a specific example of the van der Waals (vdW) mean-field model of interacting particles~\cite{Vovchenko:2015vxa}.
The model contains a first-order liquid-gas phase transition and a CP.
Scaled variance and kurtosis of fluctuations inside the subsystem as functions of $\alpha$ are shown in Fig.~(\ref{fig-1}).

We see that observed fluctuations strongly depend on $\alpha$, which displays the influence of global particle number conservation.
One also sees that for a realistic number of baryon charge $B=400$ encountered in HICS, the analytic formulas give a good approximation of the numeric results even for higher order cumulants, and even in the vicinity of the CP ($n=n_{crit}$, $T=1.2~T_{crit}$) where the correlation length is large.
\footnote{Also, SAM formulas were recently tested in a similar setting for molecular dynamics simulations of the classical Lennard-Jones fluid, where similar results were obtained~\cite{Kuznietsov:2022pcn}.}
\footnote{The present study does not consider fluctuations in the region below spinodals of the first-order phase transition. In Ref.~\cite{Poberezhnyuk:2020cen}, fluctuations in GCE below spinodals were addressed with a method similar to SAM. The subensemble results for the mixed phase region are in preparation~\cite{Savchuk}. 
}

The subensemble acceptance formulas were used to make lattice QCD-based predictions for baryon number fluctuations in HICs at LHC energies~\cite{Vovchenko:2020tsr}.
The rapidity acceptance window of unity, $\Delta Y_{acc}\approx 1$, was found to be a sweet spot where at the same time, the measurements of fluctuations are sensitive to the equation of state of QCD matter, and finite size effects are small.

The SAM formulas have been generalized to the presence of an arbitrary number of independent conserved charges~\cite{Vovchenko:2020gne}.
As an example, here the 4th order cumulant of baryon number fluctuations inside the subsystem is presented in the case when there are two conserved charges (baryon number $B$ and electric charge $Q$):
\eq{\label{eq:kappa4Btwo}
\kappa_4[B_1] & = \alpha VT^3 \, \beta \, \left[ 
 \, (1-3\alpha \beta) \, \chi_4^B
- 3 \, \alpha \, \beta \, \frac{ (\chi_3^B)^2 \chi_2^Q - 2 \chi_{21}^{BQ} \chi_{11}^{BQ} \chi_3^B + (\chi_{21}^{BQ})^2 \chi_2^B }{\chi_2^B \chi_2^Q - (\chi_{11}^{BQ})^2}
\right].
}

Formulas are modified compared to the case of a single conserved charge~\eqref{eq:kappa1B}, namely, the mixed (off-diagonal) grand-canonical susceptibilities $\chi^{BQ}$ appear.
For the SAM formulas in case of arbitrary order of cumulant and number of conserved charges as well as for non-conserved quantities such as, for example, net proton number, see Ref.~\cite{Vovchenko:2020gne}.

We show~\cite{Vovchenko:2020gne} that $\alpha$-dependence cancels out in {\it (i)} any ratio of second order cumulants, {\it (ii)} any ratio of third order cumulants {\it (iii)} ratios of mixed second order cumulants involving non-conserved quantities and conserved charge, {\it (iv)} ratio $\Sigma/\Delta$ involving so-called strongly intensive fluctuation measures~\cite{Gorenstein:2011vq} $\Sigma$ and $\Delta$.
For example, 
\eq{
\frac{\kappa_2^Q}{\kappa_2^B}=\frac{\chi_2^Q}{\chi_2^B},~~~
\frac{\kappa_3^Q}{\kappa_3^B}=\frac{\chi_3^Q}{\chi_3^B},~~~
\frac{\kappa^{BQ}}{\kappa_2^{S}}=\frac{\chi^{BQ}}{\chi_2^S},~~~
\frac{\kappa_{p\hat{Q}_j}}{\kappa_{\hat{Q}_i\hat{Q}_j}} = \frac{\chi_{p\hat{Q}_j}}{\chi_{\hat{Q}_i\hat{Q}_j}},~~~
\frac{\kappa_{p\hat{Q}_j}}{\kappa_{p\hat{Q}_i}} = \frac{\chi_{p\hat{Q}_j}}{\chi_{p\hat{Q}_i}},~~~
\frac{\kappa_{p\hat{Q}_j}}{\kappa_{k\hat{Q}_i}} = \frac{\chi_{p\hat{Q}_j}}{\chi_{k\hat{Q}_i}}.
}
Here $S$ is strangeness, $p$ and $k$ denote respectively net proton and net kaon numbers, and
$\hat{Q}_i$, $\hat{Q}_j$ are arbitrary conserved charges.
These ensemble-independent fluctuation measures are not sensitive to global conservation and are thus particularly convenient for experimental measurements.

\section{Summary}
\label{summary}

The subensemble acceptance method provides general formulas to correct cumulants of distributions in heavy-ion collisions for the global conservation of all QCD charges.
 Explicit expressions for all diagonal and off-diagonal cumulants measured in the subsystem of the thermal system that relate them to the grand canonical susceptibilities are obtained for an arbitrary equation of state with an arbitrary number of different conserved charges. 
Presented fluctuation measures are insensitive to global conservation and, thus, can be particularly convenient for experimental measurement.
The method can be extended for application at intermediate collision energies and supplemented with formulas for higher-order fluctuations of non-conserved quantities.

\end{document}